\documentclass[journal,letterpaper]{IEEEtran_nssmic}

\usepackage{cite}      

\usepackage{graphicx}  
\begin{document}
%
\title{D0 Silicon Microstrip Tracker}
%
%
\author{Sergey~Burdin~for~D0~Collaboration
\thanks{Presented at  Nuclear Science Symposium \& Medical Imaging Conference}%
\thanks{S. Burdin (burdin@fnal.gov) is with Fermilab.}}
\maketitle

\begin{abstract}
The D0 Run II silicon microstrip tracker (SMT) has
3 square meters of Si area. There are 792,576 channels 
read out by 6192 SVXIIe chips on 912 read out modules. 
The SMT provides track and vertex reconstruction capabilities 
over the full pseudorapidity coverage of the D0 detector. 
The full detector has been running successfully since April 2002. 
 
 This presentation covers the experience in commissioning 
and operating, the recent electronics upgrade which improved 
stability of the SMT and estimates of the radiation damage.
\end{abstract}


\section{Introduction}
\begin{figure}[h]
\centering
\includegraphics[width=2.5in]{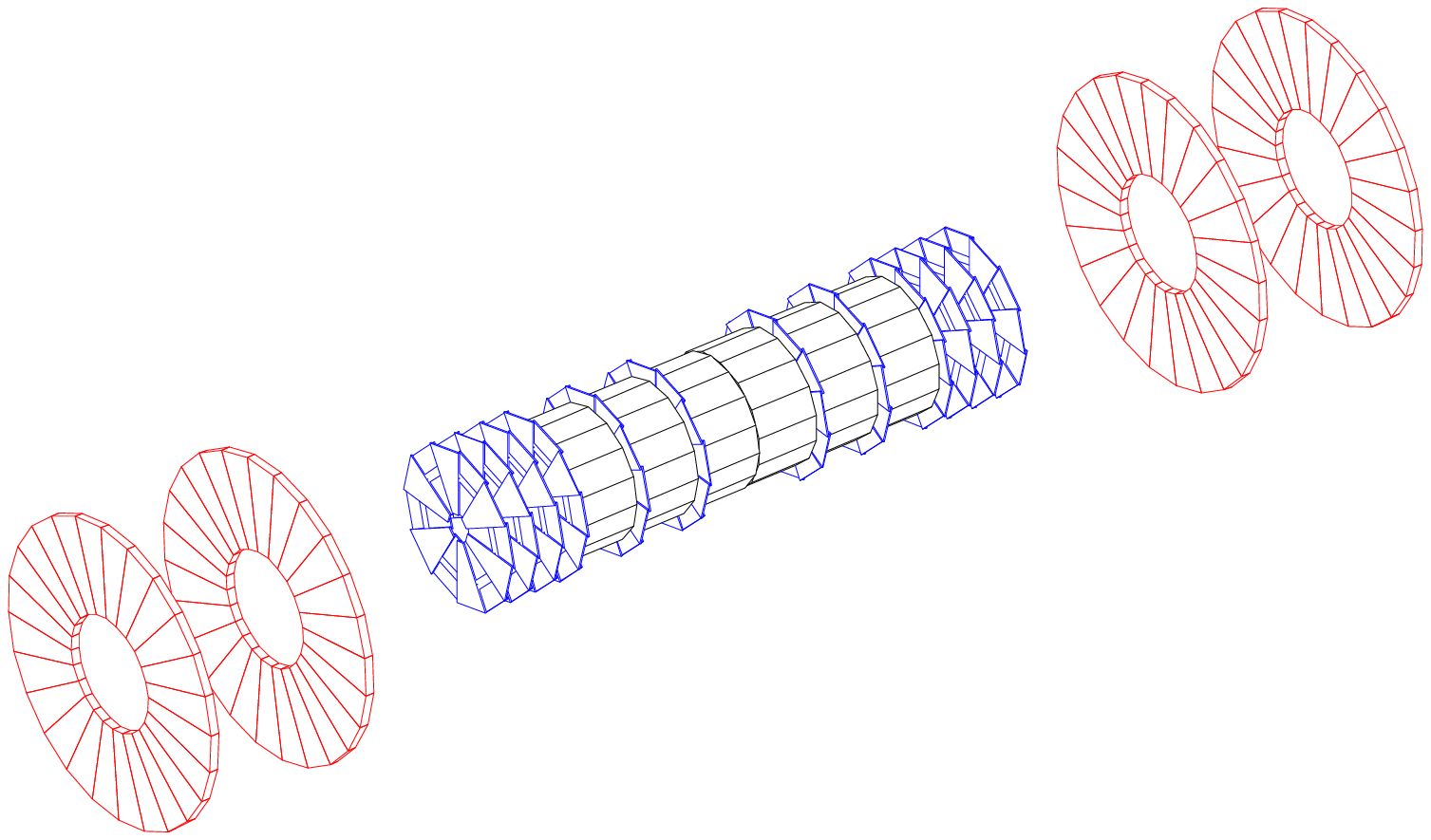}
\caption{Isometric view of the D0 silicon tracker}
\label{fig:smt_iso}
\end{figure}

%
%
%
%
\PARstart{T}{he} 
Silicon Microstrip Tracker (SMT) together with the Central Fiber Tracker (CFT) 
provide tracking and vertexing at the D0 Detector (Tevatron, Fermilab).  
The long interaction region and large pseudorapidity acceptance ($|\eta | < 3$) 
led to a hybrid design of the SMT shown in Fig.~\ref{fig:smt_iso}, 
with barrel detectors measuring primarily the $r$-$\phi$ coordinate 
and disk detectors which measure $r$-$z$ as well as $r$-$\phi$. 
Thus tracks for high $\eta$ particles are reconstructed by the disks, 
and tracks of particles at small values of $\eta$ are measured in 
the barrels. The SMT information is heavily used for many D0 analyses. 
It is crucial for selection of 
the events with t- and b-quarks, as well as for $e/\gamma$ separation.

The SMT consists of six barrels which are 12 cm long and have 72 ladders
arranged in four layers. The two outer barrels have single 
sided and double sided $2^\circ$~stereo ladders. The four inner 
barrels have double sided double metal (DSDM) $90^\circ$~stereo and double 
sided (DS) $2^\circ$~stereo  ladders. The ladders are mounted between two 
precision machined Beryllium bulkheads. 

Each barrel is capped with a disk of wedge detectors, called the ``F-disks''.
The F-disks are comprised of twelve wedges made of double sided silicon wafers
with trapezoidal shape. The stereo angle of the F-wedges is $30^\circ$. To
provide further coverage at intermediate $\eta$, the central silicon
system is completed with a set of three F-disks on each side of the 
barrel. In the far forward and backward region two large 
diameter ``H-disks'' provide tracking at high  $\eta$. 
The H-disks are made of 24 pairs of single sided detectors glued back 
to back giving a stereo angle of $15^\circ$. 

\section{Silicon Microstrip Tracker Readout}
The silicon detectors are read out using the SVXIIe chip, which 
is fa\-bri\-ca\-ted in the UTMC radiation hard 1.2~$\mu$m CMOS technology. 
Each chip consists of 128 channels, each including a preamplifier, a 32 cell 
deep analog pipeline and an 8 bit ADC. It features 53 MHz readout speed,
sparsification, downloadable ADC ramp, pedestal and bandwidth setting. 
The SVXIIe chips and associated circuitry are 
mounted on a double-sided, 0.2 mm pitch, kapton flex circuit, the so 
called High Density Interconnect, or HDI. The HDI is laminated onto a 
300 $\mu$m thick Beryllium substrate and glued to the silicon sensor. 
In case of double-sided silicon ladders, the HDI is wrapped around one silicon 
edge to serve both ladder surfaces. 
The total number of HDIs in the SMT is 912.

\begin{figure*}[htb]
\centering
\includegraphics[width=4.9in]{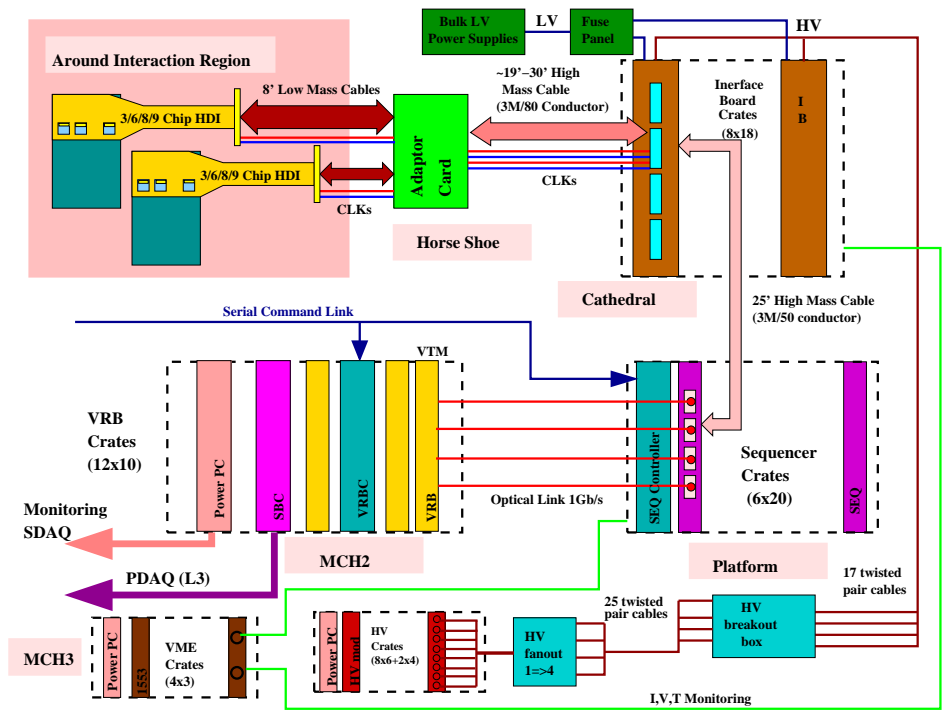}
\caption{SMT readout.}
\label{fig:readout}
\end{figure*}

Fig.~\ref{fig:readout} shows a sketch of the SMT readout setup.
The HDIs are connected through 2.5m long kapton flex cables, Adaptor 
Cards (ACs) and 10m long pleated foil cables to Interface Boards (IBs).
The ACs are located on the face of the Central Calorimeter. 
The IBs supply and monitor power to the 
SVXII chips, distribute bias voltage to the sensors and refresh data and
control signals traveling between the HDIs and the Sequencers. The Sequencers 
control the operation of the chips and convert their data into optical 
signals carried over 1Gb/s optical links to VME Readout Buffer
boards. Data is read out from the chips, transfered to the VRBs through 
the Sequencers whenever a Level-1 accept is issued and held  pending a 
Level-2 trigger decision. 
\begin{figure*}[htb]
\centering
\includegraphics[width=4.9in]{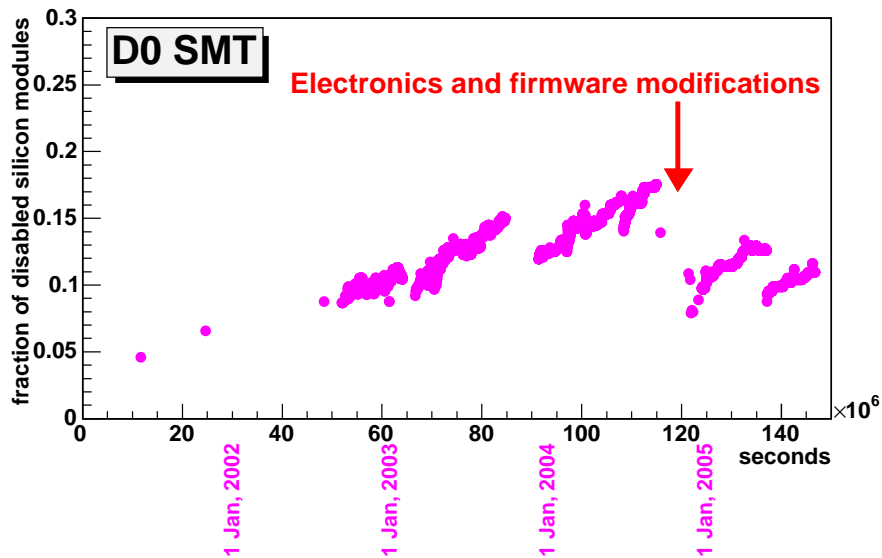}
\caption{Fraction of disabled silicon modules vs. time.}
\label{fig:disabledHDIs}
\end{figure*}

  The SMT has been fully functional since April 2002. The main operational goal 
is to keep high efficiency with low detector downtime. 
This goal is compromised by unstable
HDIs: 46\% of all HDIs were deactivated at least 
once due to readout problems. 
In most cases these HDIs can eventually be read out. 
The debugging process is 
complicated because there is no access to the upstream 
electronics starting from 
the Sequencers during Tevatron operation; 
only during shutdown one can access 
the electronics and cables up to the Adaptor Cards.

  Before the Fall 2004 shutdown 160 HDIs were disabled and 80 of them were 
declared ``dead'' due to confirmed problems in the unaccessible area. 
During the shutdown two approaches were used 
to fix the unstable HDIs. 
The traditional approach was to check the 
accessible electronic chains. In addition, special changes 
in the Sequencer firmware were
implemented to provide more stable readout. 

  Many HDI failure modes lead to loss of a proper end of readout signal from 
the HDI, thus stopping data taking.
To address
this issue, special Sequencer firmware has been developed, 
which aborts the readout if there is no activity on the signal lines for a 
certain amount of time. 
These changes and more stable procedures for 
the chip parameter download allowed us to enable many of the unstable 
HDIs. Immediately after shutdown only 80 HDIs ($\sim 10\%$ from the total number) 
were disabled (see Fig.~\ref{fig:disabledHDIs}). In Spring 2005 this 
number had grown to 120 HDIs. 
Then another firmware modification was introduced, applying a 
readout abort if the 
elapsed time exceeds the maximum time needed 
for full HDI readout. This firmware upgrade again brought 
the number of disabled silicon modules down to $\sim 10\%$. 
Taking into account HDIs with partial readout, the overall
fraction of working SVX chips is close to $83\%$.

The electronics upgrade has not only improved the silicon 
tracker efficiency 
but also significantly decreased the downtime caused 
by the silicon readout failures.

\section{Radiation Damage}

 It is expected that the lifetime of the D0 Silicon Tracker 
will be limited by noise due to 
micro-discharge breakdown~\cite{4077} in the inner layers. 
The micro-discharge effect depends on 
bias voltage and becomes unacceptable at approximately $150$~volts. 
Double sided silicon detectors require sufficient bias voltage to fully deplete 
the entire silicon volume. This depletion voltage changes with the radiation 
dose received by the detectors. To predict the lifetime 
of the D0 Silicon Tracker it is necessary to monitor the depletion voltage 
and the radiation dose. The inner layer of the inner four 
barrels consists of the DSDM silicon modules which 
were measured to be the most sensitive to the radiation~\cite{3962}. In addition, the 
inner layer receives the highest radiation dose. 
Therefore, our study is concentrated on 
the DSDM silicon modules in the inner layer.

\begin{figure}[htb]
\centering
\includegraphics[width=2.5in]{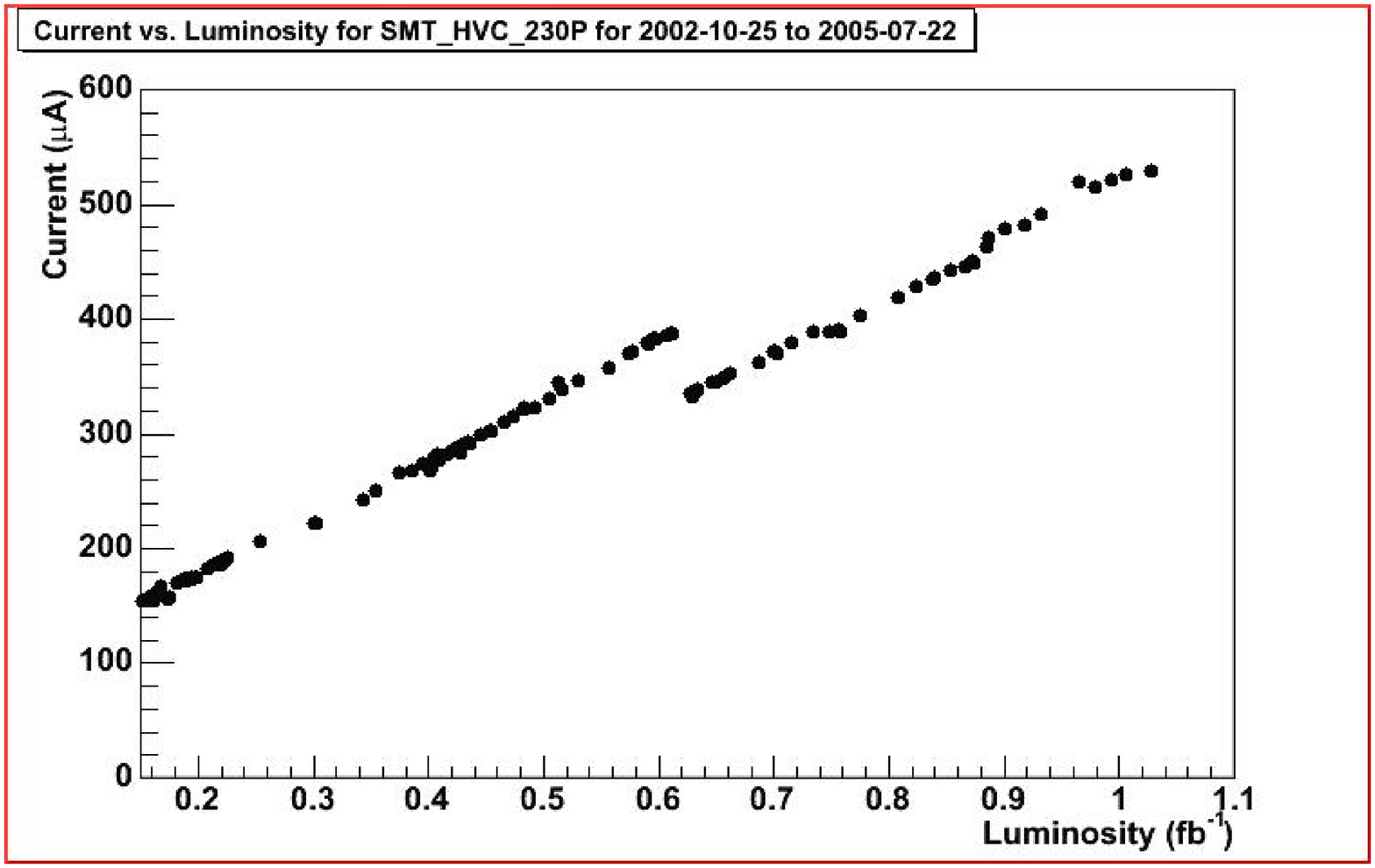}
\caption{Evolution of the leakage current normalized to $20^\circ C$ with integrated luminosity.}
\label{fig:curr}
\end{figure}
  The radiation dose received by a silicon module can be 
measured using the leakage current, 
which depends on flux according to the following formula:
\begin{equation}
I=I_{0} + \alpha \cdot \Phi \cdot V,
\end{equation}
where $I_{0}$ is the leakage current before irradiation, $\alpha$ 
is the radiation damage
coefficient, $\Phi$ is the particle flux and $V$ is the detector volume. 
The coefficient $\alpha$
depends on the particle type, temperature and time~\cite{Moll,Wunstorf}. 
The particle flux is proportional to the integrated luminosity 
delivered by the Tevatron. Therefore the dependence of the leakage current on 
the luminosity is also expected to be linear (see Fig.~\ref{fig:curr}). The drop in the 
leakage current at the luminosity $0.6$~fb$^{-1}$ is due to 
the shutdown in Fall 2004 when the silicon tracker was warmed up to
$15^\circ C$ for one month and annealing processes accelerated.
The inner silicon layer has two sub-layers: the inner sub-layer 
at radius $27.15$~mm 
and the outer sub-layer at radius $36.45$~mm. 
The radiation dose collected by the outer
sub-layer is about 1.6 times lower 
than the one at the inner sub-layer.
Measurements 
of several silicon modules in the inner layer 
gave a normalized particle flux from
\mbox{$4.0\cdot 10^{12}$} to 
\mbox{$5.3\cdot 10^{12}$~particles$/$cm$^{2}$fb$^{-1}$}, taking 
annealing parameters from~\cite{Wunstorf} and an asymptotic radiation damage constant 
of \mbox{$3\cdot 10^{-17}$~A$\cdot$cm$^{-1}$}. 
The spread 
reflects the accuracy of the method as well as possible non-uniformity 
of the radiation dose.

The silicon detector bulk material at D0 is slightly n-doped. Under
radiation, however, donor states are removed and acceptor states created,
eventually leading to a change from positive to negative space charge, i.e. the
silicon bulk goes from being n-doped to becoming p-doped. This phenomenon is
referred to as \emph{type inversion} and is confirmed by many experiments. 
It was predicted that the type inversion at the D0 silicon bulk material will 
happen when the integrated normalized particle flux will reach approximately 
$5\cdot 10^{12}$~particles$/$cm$^{2}$~\cite{3962}. The Tevatron has delivered 
slightly more than $1$~fb$^{-1}$ and, therefore, the inner layer should 
be at the type inversion point, where the depletion voltage is minimal.

  The depletion voltage can be measured using 
the dependence of the noise level on the n-side 
of a silicon module as well as the dependence 
of charge collection efficiency on the bias voltage. 

\subsection{Noise level on the n-side} 
While the silicon bulk material is n-doped, 
the noise of the n-side strips 
is large and constant at all voltages below the
depletion voltage, and reduces significantly when 
depletion is reached. 
Thus studying the n-side noise as a function of
bias voltage can be used for determination of the depletion voltage.

In a bias voltage scan 11 runs are taken, 
with the high voltage settings varied from 0 to 100\% in steps of 10\%.
The bias voltage scans were performed with and without 
beams in the Tevatron. The beam presence does not affect 
the measured depletion voltage values.

The DSDM and DS sensors show very different noise behavior 
as a function of bias voltage.
For that reason, different procedures are used for DSDM and DS devices.

The observed difference in noise behavior could come 
from the more complicated design of the DSDM 
devices~\cite{d0note3846}. 
In order to read out the n-side strips, an additional
metal layer is needed. The second metal layer is insulated from the first 
with a very thin layer of PECVD
(plasma enhanced chemical vapor deposition). 
If there is a charge build-up 
in the insulation layer, this could
change the behavior of the noise as a function of the bias voltage. 

\begin{figure}[htb]
\centering
\includegraphics[width=2.5in]{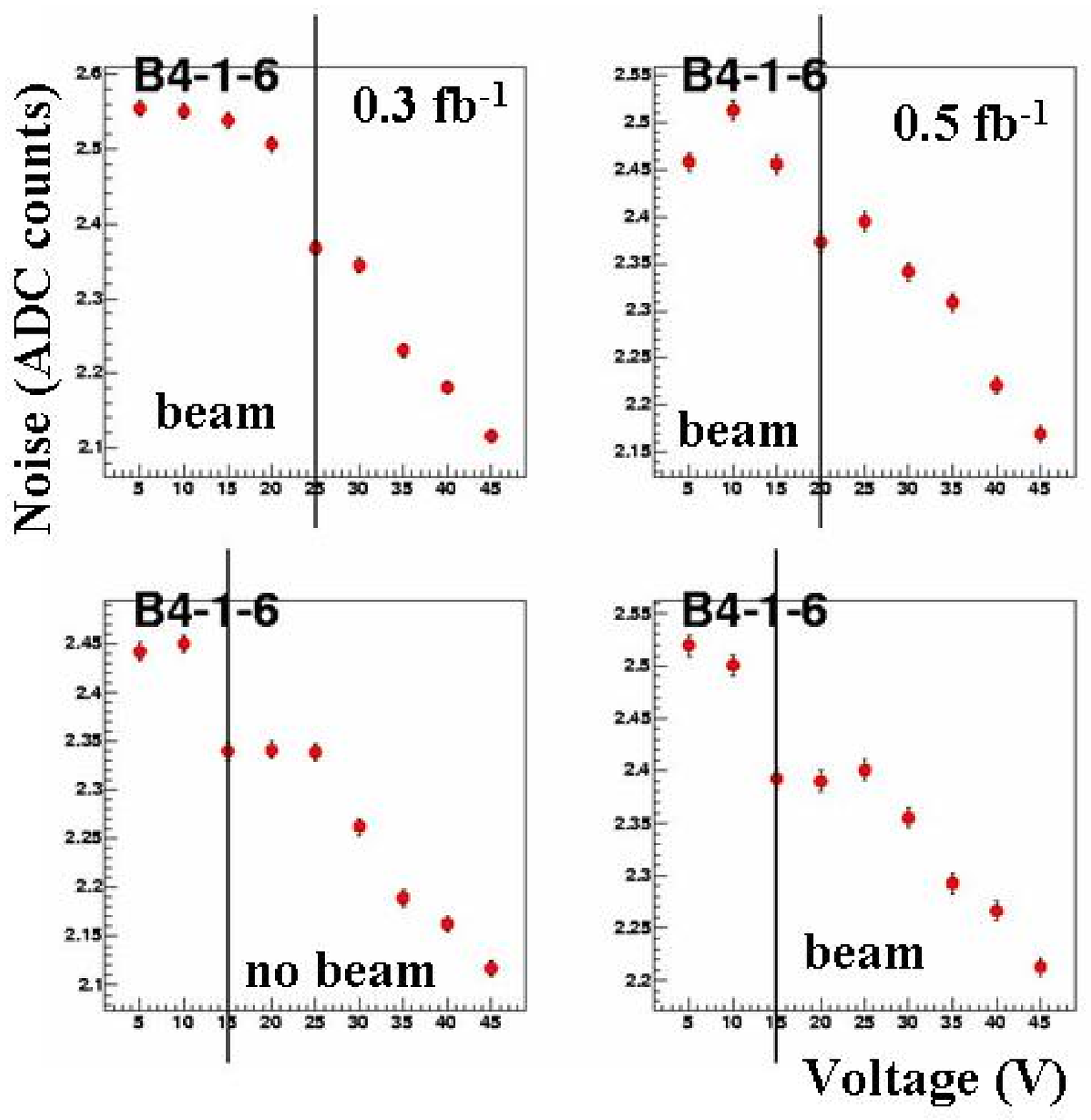}
\caption{Distribution of noise level on the n-side as a function of 
bias voltage for a double sided 
double metal silicon module installed in the D0 detector.
The bias voltage scans at different integrated luminosities are shown.}
\label{fig:dsdm_irradiated}
\end{figure}

The DSDM devices show a rather unexpected noise behavior as 
a function of bias voltage (see Fig.~\ref{fig:dsdm_irradiated}). There is no
abrupt decrease in the noise level on the n-side when the depletion voltage 
is reached. Instead, the noise is
decreasing rather monotonically with increasing bias voltage. For 
some HDIs a small kink in the noise can be
seen at a certain bias voltage, and the position of this kink is 
changing as a function of the radiation dose.
We interpret this kink as an indication of the depletion voltage. 

\begin{figure}[htb]
\centering
\includegraphics[width=2.5in]{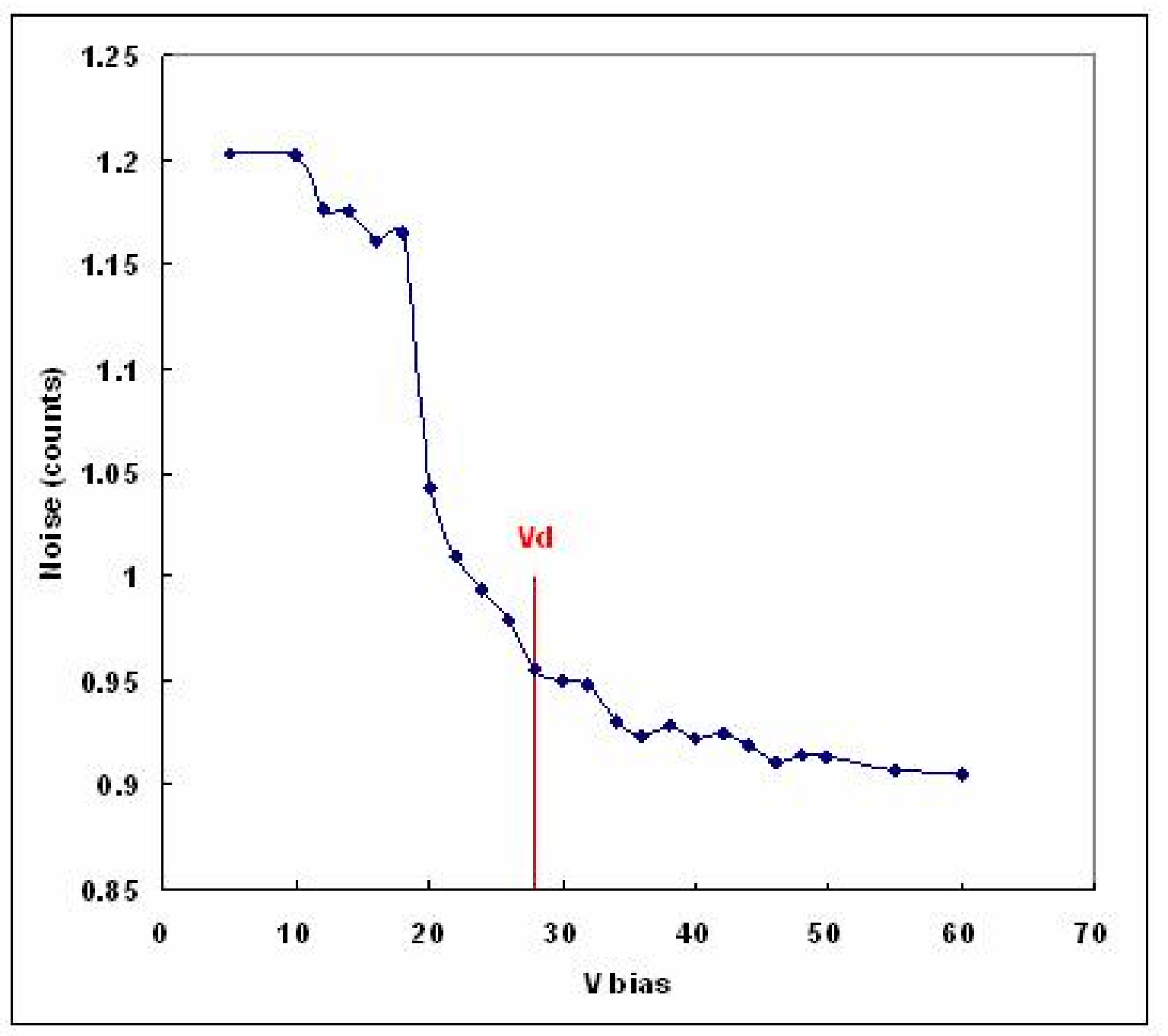}
\caption{Distribution of noise level on the n-side as a function of 
bias voltage for a non-irradiated
double sided double metal test module.}
\label{fig:dsdm_nonirradiated}
\end{figure}

The n-side noise was measured at a non-irradiated test DSDM module 
(see Fig.~\ref{fig:dsdm_nonirradiated}) and abnormal behavior 
was not observed. This measurement indicates that the abnormal 
noise behavior is caused by the irradiation, and the comparison with the 
DS modules shows that the radiation changes some properties of the PECVD 
layer.

\begin{figure}[htb]
\centering
\includegraphics[width=2.5in]{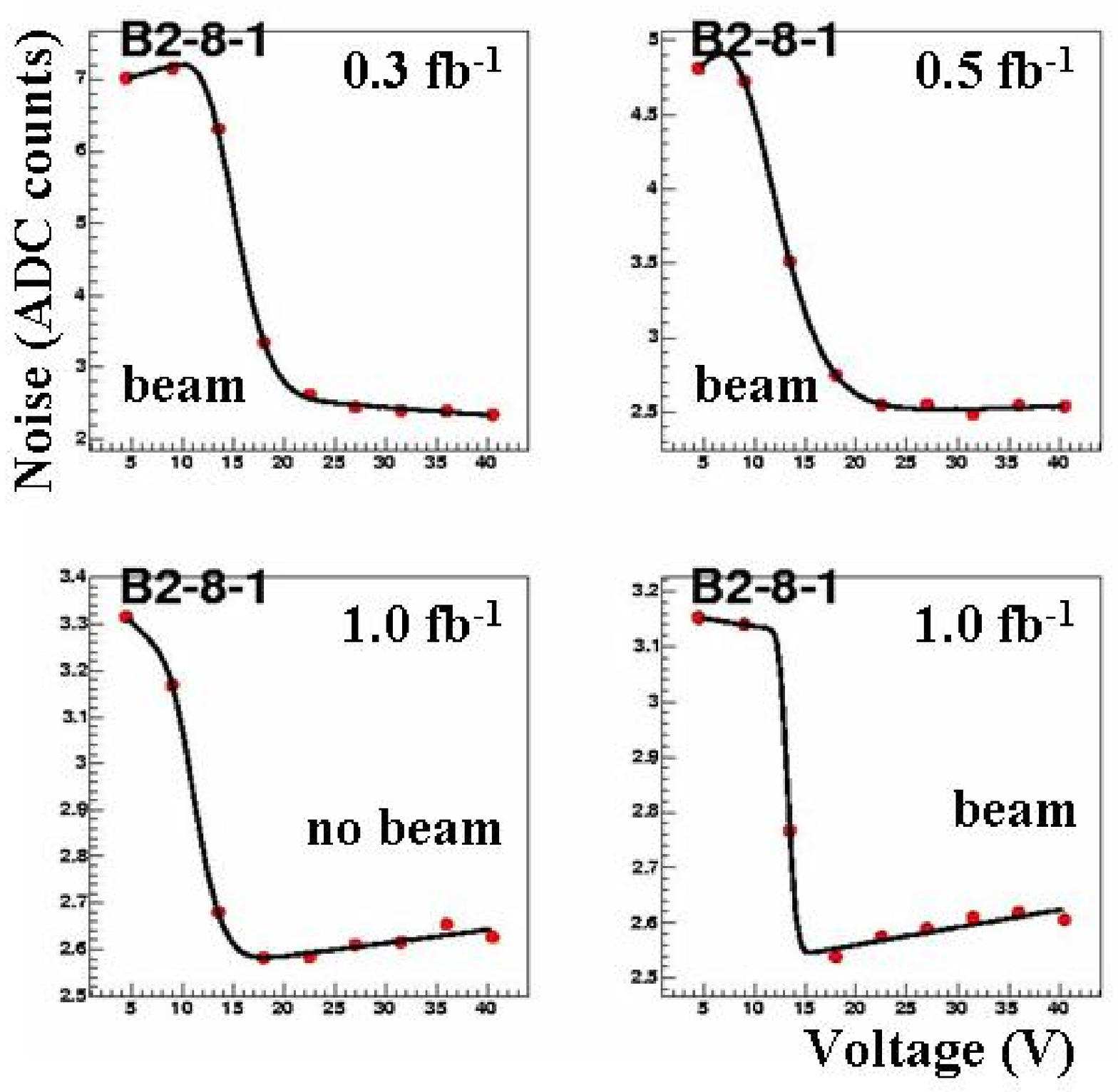}
\caption{Distribution of noise level on the n-side 
as a function of bias voltage for a double sided 
silicon module installed in the D0 detector.
The bias voltage scans at different integrated luminosities are shown.}
\label{fig:ds_irradiated}
\end{figure}

Most of the DS devices show the expected noise behavior 
as a function of bias voltage (see Fig.~\ref{fig:ds_irradiated}). 
At low voltages the
noise is large, and rapidly decreases to a stable and 
lower level as soon as the bias voltage
reaches the depletion voltage. There is some 
indication of noise increase at bias voltages higher than the depletion 
voltage at higher radiation doses.

\subsection{Charge collection efficiency} 
Another method for determination 
of the depletion voltage uses the dependence of 
charge collection efficiency on bias voltage. 
For the silicon detectors with n-doped bulk material, the 
charge collection efficiency measured on the p-side increases 
with bias voltage and reaches its maximum at full depletion.
Though 
there could be some additional small increase after that, for example, 
due to change in the charge collection time.

\begin{figure}[htb]
\centering
\includegraphics[width=2.5in]{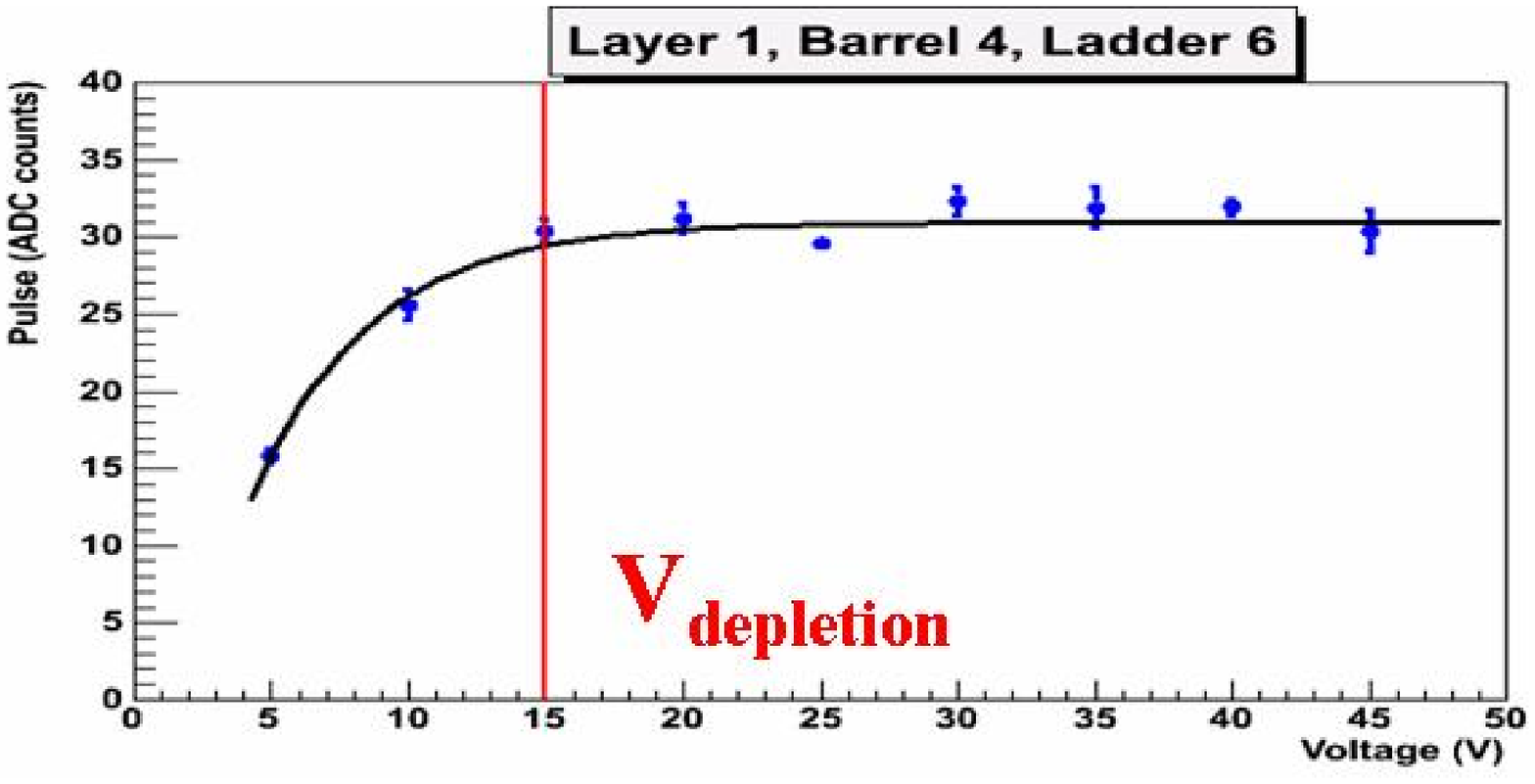}
\caption{Dependence of the charge collection efficiency 
on bias voltage for the double sided 
double metal silicon module installed in the D0 detector.}
\label{fig:dsdm_cce}
\end{figure}

This method requires bias voltage scans with tracks in the detector. 
A special algorithm has been developed for cluster reconstruction on 
the ladder under study in the vicinity of the expected track position. 
To determine the charge collection efficiency 
the cluster charge is measured. 
The bias voltage where the 
charge collection efficiency reaches 95\% of its asymptotic value has
been chosen as the depletion voltage (see Fig.~\ref{fig:dsdm_cce}). 

\begin{figure}[htb]
\centering
\includegraphics[width=2.5in]{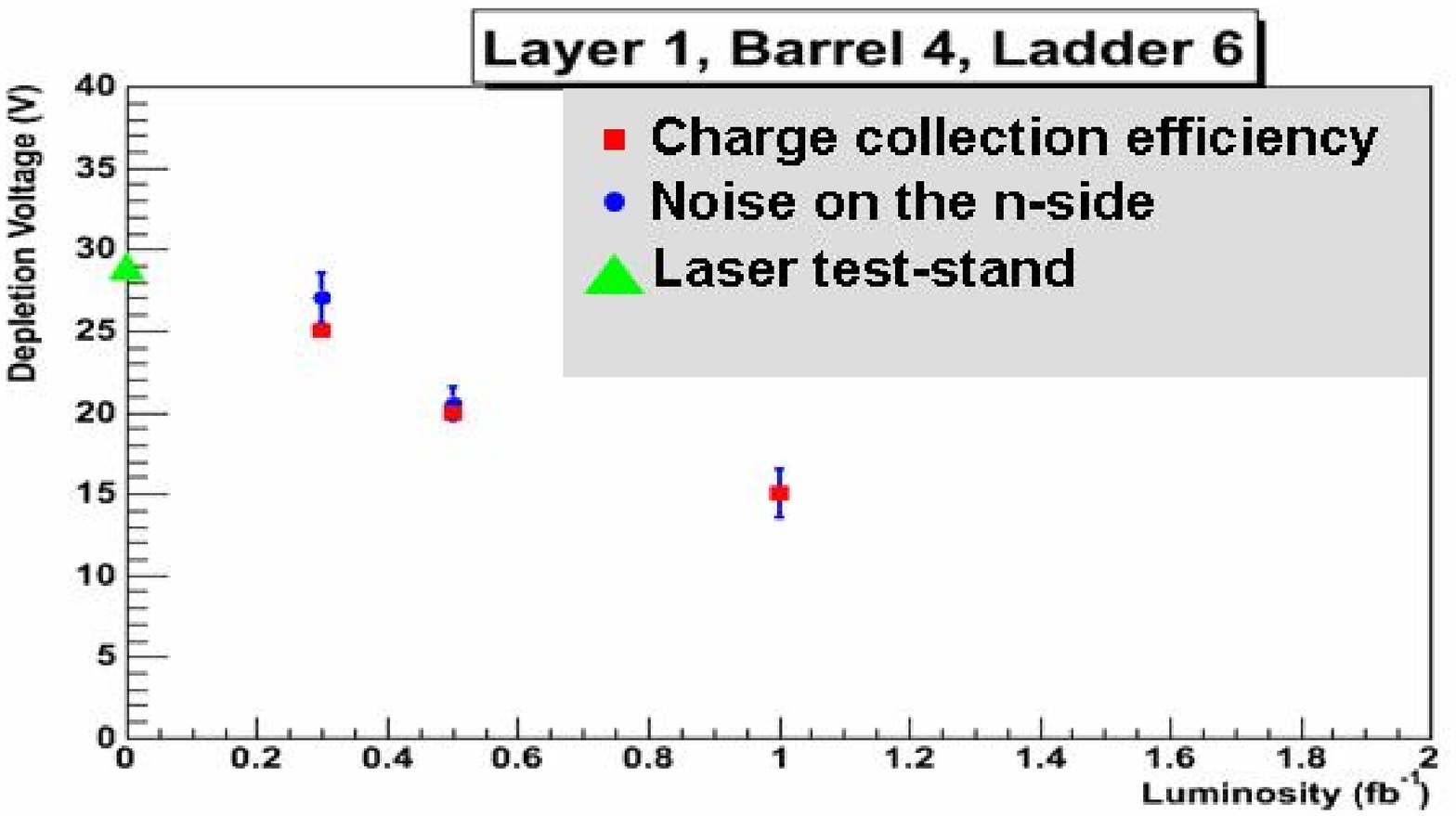}
\caption{Comparison of depletion voltages determined 
by two methods.}
\label{fig:dvcomp}
\end{figure}

The depletion voltage measured using 
the charge collection efficiency shows agreement with 
the noise measurements (see Fig.~\ref{fig:dvcomp}). 
The silicon modules also have been measured at a laser test-stand 
before they have been installed in the D0 detector.
The corresponding initial depletion voltage is also shown in Fig.~\ref{fig:dvcomp}.
\begin{figure*}[htb]
\centering
\includegraphics[width=4.5in]{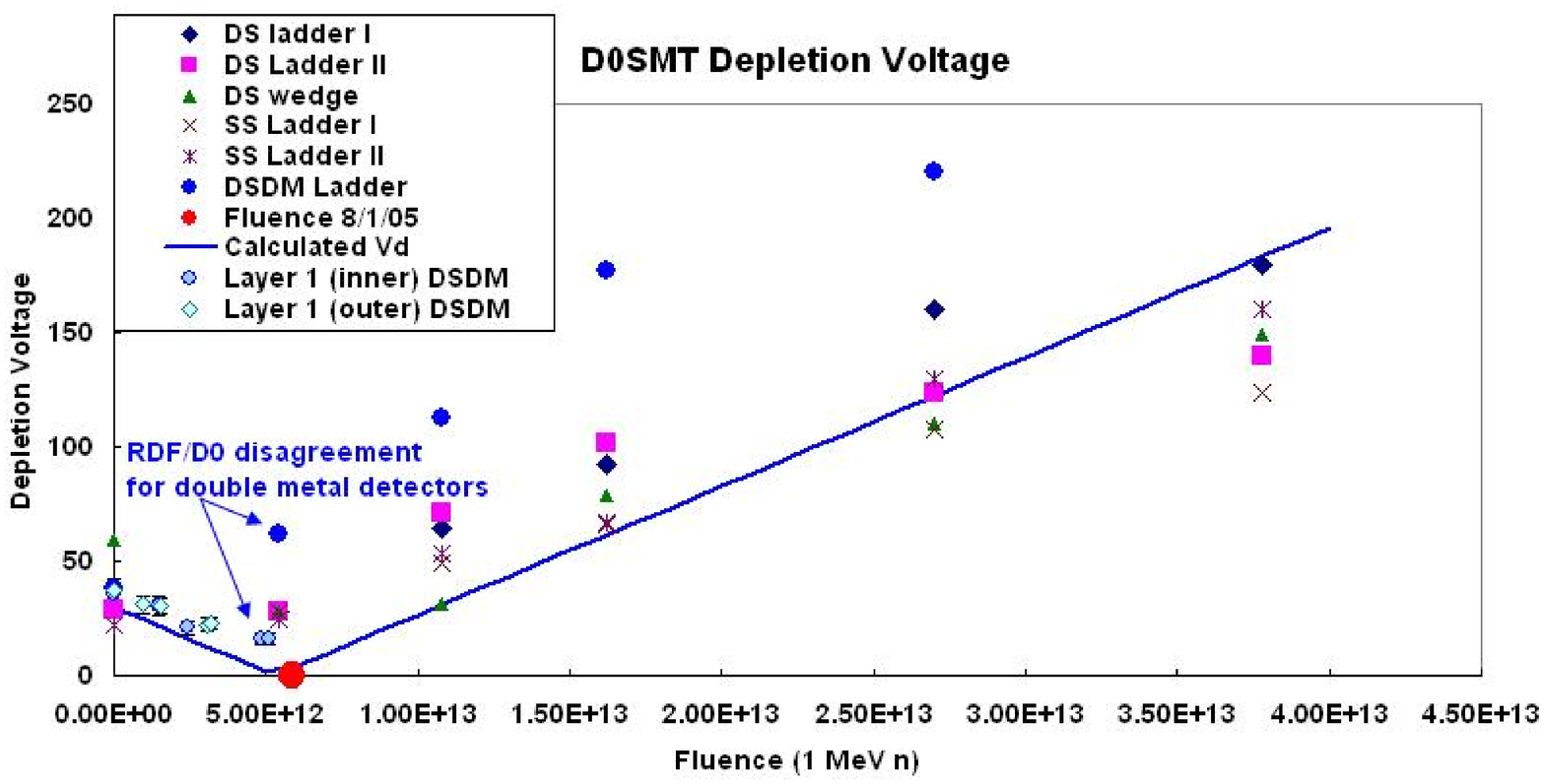}
\caption{Depletion voltage for the silicon 
modules installed in the D0 Detector and irradiated at the 
Radiation Damage Facility for different radiation doses. 
The line represents a theory prediction according to~\cite{Moll}.
}
\label{fig:booster_comparison}
\end{figure*}

\subsection{Lifetime estimate}

 The measurements of the radiation dose and 
depletion voltage allow us to compare the 
behavior of the silicon sensors installed 
in the D0  Silicon Tracker with the ones 
irradiated at the Radiation Damage 
Facility (RDF) at Fermilab. 
The comparison in Fig.~\ref{fig:booster_comparison} 
shows that the 
DSDM silicon modules irradiated at RDF and 
those installed in the D0  Detector have a 
different dependence of the depletion voltage 
on the radiation dose. The depletion
voltage for the DSDM irradiated sensors increases 
much faster than for the  single metal double 
sided silicon detectors tested at RDF. 
From our measurement, the initial behavior of depletion voltage for 
the DSDM silicon modules installed in the D0  
Detector is similar to the  behavior
of the single metal double sided modules and 
is well described by standard parameterizations~\cite{Moll}. 
The discrepancy is probably due to slow annealing 
of charge trapped in the insulator layer between 
metal layers which was not measured in the RDF studies. 
Assuming this ``normal'' behavior continues, the depletion 
voltage for the DSDM silicon sensors at the inner layer 
will reach values around $150$~volts at a delivered 
luminosity of \mbox{$5$--$7$~fb$^{-1}$}.

\section{Conclusion}
The D0 Silicon Tracker has been working successfully 
since April 2002. 
Recent electronics and firmware upgrades improved 
its efficiency and stability. Depletion voltage 
measurements for the inner DSDM silicon modules 
show the expected 
radiation damage of the silicon bulk material. The 
thin layer of
plasma enhanced chemical vapor deposition, 
which is used for 
insulation of two metal layers in the DSDM sensors, 
is affected by the radiation as well. This causes 
abnormal n-side noise behavior and probably biases the 
depletion voltage measurements for the DSDM silicon modules 
irradiated at the Radiation Damage Facility.
If the silicon tracker lifetime is limited
by only the depletion voltage and the breakdown does not 
happen until bias voltages $\sim 150$~V, then the D0 silicon
tracker will stay operational up to 
delivered luminosities of \mbox{$5$--$7$~fb$^{-1}$}.


%
%


\section*{Acknowledgment}
This work is presented on behalf of the D0 Collaboration.
The author would like to thank John Omotani (University of Cambridge, UK), who has made
significant contributions to the leakage current study while a summer student at 
the D0 experiment.



%




\end{document}